\documentclass[12pt,letterpaper]{article}
\usepackage{multirow}
\usepackage{slashed}

\title{The Schwinger and Chiral Schwinger Models in a Non-perturbative Spectral Regularization}
\author{L. F. Eleot{\'e}rio, G. A. Oliveira, \\ E. W. Dias, H. Caldas and A. L. Mota\footnote{motaal@ufsj.edu.br}\\
\small{Departamento de Ci\^{e}ncias Naturais,}\\
 \small{Universidade Federal de S\~{a}o Jo\~{a}o del Rei,}\\
 \small{C.P. 110,  CEP 36301-160, S\~ao Jo\~ao del Rei, Brazil}}

\date{}

\usepackage{epsfig}
\usepackage{epsf}

\def \be{\begin{equation}}
\def \ee{\end{equation}}

\begin{document}
\maketitle

\vspace{10pt}

\begin{abstract}

We investigate the employment of a non-perturbative regularization scheme -- the spectral regularization, which is based on the gauge technique, previously implemented in the context of chiral quark models -- in the study of the gauge symmetry preservation within the Schwinger model and violation in the chiral Schwinger model. We show that the spectral regularization provides mathematical consistent and ambiguity free solutions for the two-point functions of both Schwinger and chiral Schwinger models in exact (1+1) dimensions, correctly displaying the gauge invariance in the Schwinger model and the axial anomaly in the chiral Schwinger model. The employment of the spectral regularization avoids any dependence on ambiguous amplitudes and/or unconventional $\gamma^5$ algebra. Our results reinforce the strength of the spectral regularization as a mathematical consistent, divergence free, and ambiguity free regularization scheme that correctly implements symmetry conservation or violation in each case.

\end{abstract}

\section{Introduction}

The Schwinger Model \cite{Schwinger1951} -- the quantum electrodynamics of massless fermions in (1+1) dimensions -- is a paradigm to the study of quantum gauge field theories. It is an exactly solvable model \cite{Niemi1986,Ruiz1986,Karsch1995,Linares2001}  that presents similar features with realistic models such as the 4D quantum electrodynamics (QED) and, mainly, quantum chromodynamics (QCD). It exhibits fermions confinement \cite{Coleman1975} and, in its chiral version \cite{Manton1985}, chiral symmetry breaking and a chiral anomaly, similar to the QCD. The Schwinger model, and other closely related (1+1)D fermionic models, have been recently applied as an effective model for condensed matter 1D fermionic systems such as graphene wires, polymeric chains, and 2D topological materials 
\cite{Faccioli2009,Caldas2009,Caldas2009_2,Bazzanella2010,Caldas2010,Caldas2011,Magnifico2019,Tirrito2022}.

In its perturbative approach \cite{Adam1998}, the Schwinger model is finite, in the sense that no divergences are left at the end of the calculation of probability amplitudes. However, despite its finiteness, the model suffers from the presence of ambiguous quantities \cite{Jackiw2000}, i.e., regularization dependent quantities that result from a mathematical indeterminacy in the connection limit \cite{Morais2011}. As a renormalizable model, the presence of an ambiguity does not represent any obstacle to the physical interpretation of the model, since it can be removed by the renormalization procedure, with the introduction of the proper invariant counter terms. However, being finite, no counter terms are needed to remove divergences of the model, only the ambiguous terms. So, it is natural to ask whether the symmetries of the model can be realized despite (or as a consequence of) the presence of the ambiguous term. This issue was properly addressed in ref. \cite{Jackiw2000}, in which this ambiguity is explored to ensure the preservation of a fundamental expected symmetry of the model -- the gauge invariance, expressed by the transversality of the vacuum polarization tensor (VPT). However, by doing so, a formal property of the vacuum polarization tensor had to be abandoned.

In fact, when explicitly evaluated in exactly (1+1) dimensions, the trace of the VPT in the Schwinger model is null. It can be verified by an explicit regularization independent calculation, and it implies that the ambiguous term present in the VPT should also be null. However, by accepting this formal result, gauge invariance is not satisfied. To ensure gauge invariance, the nullity of the ambiguous term and, in consequence, the tracelesness of the VPT has to be abandoned. In what follows we will refer to the preservation of the value of the trace of the Vacuum Polarization tensor in all the intermediary steps of the amplitude calculation, from the beginning to the final result, as the trace identity. As pointed out in ref. \cite{Jackiw2000}, to preserve gauge invariance in the Schwinger Model, the trace identity must be violated.

In the context of the Schwinger or Chiral Schwinger model, this is not a puzzle at all, since the  introduction of the proper counter terms eliminates any ambiguities or other illnesses introduced by the ill-defined product of local distributions in the perturbative evaluation of quantum amplitudes \cite{Zimmermann1969,Bonneau2001,Bonneau2007}, rendering physical meaning for the renormalizable models. In this sense, whether or not the VPT trace identity is preserved is not relevant, although it can be used as a test on the search for mathematical consistent and symmetry-preserving regularization schemes \cite{Mota1998,Scarpelli2001}. 

There is a class of models, however, where both these features (mathematical consistency and symmetry-preserving results) have significant relevance -- the non-renormalizable chiral quark models \cite{Vogl1991,Klevansky1992,Volkov1993,Hatsuda1994,Suzuki1998,Cheng1998}. In the framework of these models, the regularization procedures are part of them, since neither the divergent nor the ambiguous intermediary terms that arise on the computation of the amplitudes can be eliminated by a renormalization procedure. In addition, the presence of the $\gamma_5$ Dirac matrix introduces other difficulties relative to the employment of regularization procedures in non-integer dimensions (dimensional regularization \cite{tHooft1972}, for instance), or can even result in non-conventional algebraic relations involving the Dirac $\gamma_5$ matrices -- an issue also present in the Chiral Schwinger Model, as shown in ref. \cite{Viglioni2016}. Within these models, mathematical consistency, in addition to symmetry-preserving results, seems to be essential ingredients for a suitable regularization scheme, and, in consequence, for the models results. A very interesting and promising regularization technique is the non-perturbative spectral regularization, developed in the context of the so-called Spectral Quark Model \cite{Arriola2003}.

In Quantum Field Theory, a traditional attempt to improve the results and avoid divergences in the evaluation of QFT amplitudes corresponds to the employment of non-perturbative calculations, corresponding, at least formally, to resumming some specific set of diagrams in different orders in the perturbative expansion. For example, the so-called gauge technique \cite{Salam1964,Delbourgo1977,Delbourgo1979,Haeri1988} corresponds to finding some particular representation for the vertex functions of gauge models, as a functional of the fermion propagator, that solves the Ward-Takahashi identities \cite{Delbourgo1979}. If one considers the exact propagator for the theory (the Lehmann-K{\"a}ll{\'e}n propagator \cite{Lehmann}) the two-point propagators can be expressed using the spectral density of energy of the fermion, presumably computed from the fermion self-energy. So, the particular solution of the vertex function will also be expressed in terms of this spectral density \cite{Delbourgo1979}. One expects the resulting amplitudes computed from these vertex solutions to be well-defined (without divergences and ambiguities), up to transverse terms that cannot be defined by the Ward-Takahashi identities, and this feature can be used to impose some expected conditions to be satisfied by the unknown spectral density. This prescription was successfully applied to the low energy phenomenology of the Quantum Chromodynamics (QCD) \cite{Sauli2020} and, particularly, in the context of the Spectral Quark Model \cite{Arriola2003,Arriola2007,Broniowski2007,Broniowski2008,Ferreira2009,Broniowski2020}. The results of this spectral regularization of a quark model are consistent with all the observed light mesons phenomenology of QCD, and an explicit form of the spectral density for the model was provided \cite{Arriola2003}. Another interesting aspect of this solution is that the chiral anomaly of QCD can be correctly obtained \cite{Ferreira2009}, provided that the effective coupling between pseudo-scalar mesons and quarks shows a specific dependence with the spectral mass (the parameter of the spectral distribution to be integrated on), reproducing the Goldberger-Treiman relation \cite{Goldberger1958}, reinforcing thus the relevant role of the coupling constant dependence on the spectral mass in reproducing the expected symmetries of the model. The success of the spectral regularization in generating symmetry-preserving results, also eliminating divergences and ambiguities in the context of chiral quark models, suggests this regularization procedure to be relevant in the study of non-renormalizable models. These results motivated us to investigate whether the spectral regularization can be employed on the Schwinger and Chiral Scwhinger model and if it is able to avoid the ambiguous results that lead to violation of the trace identity and the non-conventional $\gamma_5$ algebra.

So, in this work, we will show that, by employing the spectral regularization to the Schwinger Model, we can find the conditions for both gauge invariance and trace identity of the VPT to be preserved. Besides, we will show explicit solutions for the spectral density and the electromagnetic coupling that fulfills all symmetry requirements of the model.

This paper is organized as follows: in section 2 we present the calculation of the vacuum polarization tensor in the Schwinger model and discuss why the violation of the trace identity is necessary to preserve gauge invariance in exact (1+1)D perturbative regularizations. In section 3 we briefly review the spectral regularization, based on the gauge technique, and apply it to the Schwinger model. We also obtain the spectral conditions to be satisfied in order to preserve both the trace identity and gauge invariance. A particular solution of the spectral density and spectral coupling that fulfills the aforementioned spectral conditions is presented in section 4. In section 5 we present the computation of the vector/axial vector two-point function within the chiral Schwinger model, showing that the manifestation of the chiral anomaly can be correctly addressed in the scope of the spectral regularization without both the presence of ambiguities and problems related to the $\gamma^5$ algebra in integer dimensions or divergent amplitudes. Finally, section 6 brings our concluding remarks.

\section{The Model}\label{TheModel}

The Schwinger model is defined by the Lagrangian density
\be
\mathcal{L}= \bar{\psi} (i \slashed{\partial} + m) \psi - e_0 \bar{\psi} \gamma_\mu A^{\mu} \psi, \label{lag}
\ee
where $\psi$ and $A^{\mu}$ are the fermionic and gauge fields respectively and $e_0$ is the bare coupling constant. We have $m=0$ in the Schwinger model, this feature allows exact solutions and also local and global chiral symmetry in its chiral version. However, let us keep $m$ explicit in the following expressions and take $m=0$ when appropriate. The Dirac gamma matrices are given, in (1+1)D, by two of the Pauli matrices, i.e.,
\be
\gamma^0 = \left[ \begin{array}{cc}
0 & 1 \\ 
1 & 0
\end{array} \right],
\gamma^1 = \left[ \begin{array}{cc}
0 & -1 \\ 
1 & 0
\end{array} \right]. \label{DiracMatrices}
\ee

The metric in (1+1)D Minkowski space is given by $g^{\mu \nu} = diag(1,-1)$. The Vacuum Polarization Tensor in momentum space is computed, from Eq. (\ref{lag}), as
\be
\Pi^{\mu \nu}(q) = \int \frac{d^2 p}{(2\pi)^2} Tr\left\{ i \gamma^\mu \frac{i}{\slashed{p}-m} i \gamma^{\nu} \frac{i}{(\slashed{p} - \slashed{q})-m} \right\}. \label{TPV1}
\ee
The trace in Dirac space in exact (1+1) dimensions can be computed directly from the Dirac matrices definition, Eq. (\ref{DiracMatrices}), resulting in
\be
\Pi^{\mu \nu}(q) = \int \frac{d^2 p}{(2\pi)^2} \frac{2(p^\mu p^{\prime \nu} + p^\nu p^{\prime \mu} - p \cdot p^{\prime} g^{\mu \nu} + m^2 g^{\mu \nu})}{(p^2-m^2)(p^{\prime 2}-m^2)}, \label{TPV2}
\ee
where we have defined $p^{\prime} = p - q$. From Eq. (\ref{TPV2}) we obtain the trace in Minkowski space, $g_{\mu \nu} \Pi^{\mu \nu}$, as
\be
\Pi^{\mu}_{\mu} = 4 m^2 \int \frac{d^2 p}{(2\pi)^2} \frac{1}{(p^2-m^2)(p^{\prime 2}-m^2)}. \label{Pimumu} 
\ee
From Eq.(\ref{Pimumu}) we can see that, for $m^2 \equiv 0$,  $\Pi^{\mu}_{\mu} = 0$. Even computing Eq. (\ref{Pimumu}) for non-zero $m$ and taking the limit $m^2 \rightarrow 0$ at the end of the calculations we get a null trace for the VPT. Thus the perturbative result for the VPT of the Schwinger model in exact (1+1)D has a null trace ($\Pi^{\mu}_{\mu} = 0$). 

Continuing with the calculation, after introducing one Feynman parameter we obtain, from Eq. (\ref{TPV2}),
\begin{eqnarray}
&&\Pi^{\mu \nu} = 2 \int_0^1 dx \int \frac{d^2 p}{(2\pi)^2}  \frac{2 p^{\mu} p^{\nu} - p^2 g^{\mu \nu}}{((p-qx)^2-M^2)^2} \nonumber \\
&& + 2 \int_0^1 dx \int \frac{d^2 p}{(2\pi)^2} \frac{-p^{\mu}q^{\nu}-p^{\nu}q^{\mu}+(p \cdot q + m^2) g^{\mu \nu}}{((p-qx)^2-M^2)^2}, \label{VPT3}
\end{eqnarray}
with $M^2 = m^2 + q^2 x(x-1)$. The first integral on the r.h.s. of Eq. (\ref{VPT3}) is superficially logarithmically divergent, and will be referred, in what follows, as $\Delta^{\mu \nu}$. Due to the logarithmic degree of divergence, we can perform the shift $p-qx \rightarrow p$ on the integrand without introducing any surface terms, and then we obtain
\be
\Delta^{\mu \nu} = \int_0^1 dx \int \frac{d^2 p}{(2\pi)^2} \frac{2 p^{\mu} p^{\nu} - p^2 g^{\mu \nu}}{(p^2-M^2)^2}. \label{Deltamunu}
\ee

The second integral on the r.h.s. of Eq. (\ref{VPT3}) is finite and unambiguous, and can be easily computed. At the end of the calculations, we get
\be
\Pi^{\mu \nu} = 2 \Delta^{\mu \nu} + \left(\frac{q^{\mu} q^{\nu}}{q^2} - \frac{g^{\mu \nu}}{2} \right) \frac{i}{\pi} -  \left(\frac{q^{\mu} q^{\nu}}{q^2} - g^{\mu \nu} \right) \frac{i}{\pi} m^2 Z_0(q^2,m^2), \label{VPTfinal}
\ee
where 
\be
Z_0(q^2,m^2) = -\frac{4}{\sqrt{q^4 - 4 m^2 q^2}} \textrm{ arctanh}\left( \frac{q^2}{\sqrt{q^4 - 4 m^2 q^2}} \right). \label{Z0}
\ee

The term $\Delta^{\mu \nu}$ is independent of $M^2$ (as can be verified by checking, in an explicit calculation, that $\partial \Delta^{\mu \nu}/ \partial M^2=0$) and superficially logarithmically divergent, but results finite in any regularization scheme, although undetermined \cite{Jackiw2000,Morais2011}. For example, an explicit calculation employing a sharp cut-off on $p$ gives $\Delta^{\mu \nu} = 0$. In contrast, in dimensional regularization, replacing $d^2 p \rightarrow d^{\omega} p$ in Eq. (\ref{Deltamunu}) and taking the limit $\omega \rightarrow 2$ at the final step, gives $\Delta^{\mu \nu} = -\frac{i}{4 \pi} g^{\mu \nu}$. By replacing this result in Eq. (\ref{VPTfinal}), we get a gauge invariant Vacuum Polarization Tensor in the limit $m^2 \rightarrow 0$. In this case, the trace of the VPT in Minkowski space is non-null, but the trace identity is satisfied, since in 
$\omega$ dimensions the trace of Eq. (\ref{TPV2}) is non-null from the beginning. Thus within dimensional regularization both the trace and gauge identities are preserved.  

In this work, our interest is in the evaluation of the VPT in exact (1+1)D. In this case, the trace of the term $\Delta^{\mu \nu}$, Eq. (\ref{Deltamunu}), is also manifestly null, since the trace of $2 p^{\mu} p^{\nu} - p^2 g^{\mu \nu}$ is equal to zero in (1+1)D, i.e., $\Delta^\mu_\mu = 0$. By observing that the only Lorentz covariant tensor of rank 2 that can result from the computation of Eq. (\ref{Deltamunu}) is $\Delta^{\mu \nu} = \alpha g^{\mu \nu}$, where $\alpha$ is a scalar constant, \footnote{If we abandon covariance requirements, by adopting different independent cut-offs for the integrals in $p^0$ and $p^1$ in Eq. (\ref{Deltamunu}), for example, then an undetermined result for this quantity can be obtained \cite{Morais2011}.}, we conclude that the trace of $\Delta^{\mu \nu}$ is $\Delta^{\mu}_{\mu} = 2 \alpha = 0$, which implies in $\alpha = 0$ and, consequently, $\Delta^{\mu \nu} = 0$. 

Returning to Eq. (\ref{VPTfinal}), and observing that $\lim_{m^2 \rightarrow 0} m^2 Z_0(q^2,m^2) = 0$, we obtain 
\be
\Pi^{\mu \nu} = \left(\frac{q^{\mu} q^{\nu}}{q^2} - \frac{g^{\mu \nu}}{2} \right) \frac{i}{\pi},
\ee
and, in consequence, $\Pi^\mu_\mu = 0$ and $q_\mu \Pi^{\mu \nu} = \frac{i q^\nu}{2 \pi}$, i.e., the trace identity is preserved but gauge invariance is not observed. At this point, it is worth observing that the third term on the r.h.s. of Eq. (\ref{VPTfinal}) is gauge invariant, however it vanishes in the limit $m^2 \rightarrow 0$. 

Jackiw \cite{Jackiw2000} suggested that we must abandon the trace identity (or, to be more specific, assume $\Delta^{\mu \nu} \ne 0$) in order to preserve gauge identity, assuming a non-null $\Delta^{\mu \nu} = -\frac{i}{4 \pi} g^{\mu \nu}$ in such a way that $\Pi^{\mu \nu} = \left(\frac{q^{\mu} q^{\nu}}{q^2} - g^{\mu \nu} \right) \frac{i}{\pi}$, a gauge invariant result. This subject was already properly studied in ref. \cite{Jackiw2000}, correctly favoring the physical requirement (gauge invariance) instead of the formal requirement (trace identity).

Let us show, however, that both identities can be consistently fulfilled in a non-perturbative approach, closely related to the aforementioned gauge technique \cite{Delbourgo1979}. Let us start briefly reviewing the spectral regularization \cite{Arriola2003} and introducing our approach in the sequence.

\section{The Spectral Regularization}

The gauge technique \cite{Delbourgo1977,Delbourgo1979} corresponds to finding a particular solution of the relevant Ward-Takahashi identities of a gauge model to define the vertex functions of the model. For example, the Ward-Takahashi identity for the vector-fermion-fermion vertex with the coupling constants factorized out reads
\be
q_{\mu} \Gamma^{\mu}(p,p-q) = S(p) - S(p-q), \label{VecWT}
\ee
where $p$ and $-(p-q)$ are the momenta of the outgoing fermions (and $q$ is the momentum of the incoming gauge field), $S(p)$ is the full propagator of the fermion, in the generalized Lehmann representation \cite{Lehmann}
\be
S(p) = \int d\omega \rho(\omega) \frac{i}{\slashed{p}-\omega + i\epsilon}, \label{LehmannProp}
\ee
and $\Gamma^{\mu}(p,p-q)$ is the unamputated vector vertex. Here, $\rho(\omega)$ is the Lehmann-K\"{a}ll\'{e}n spectral density. A particular solution of Eq. (\ref{VecWT}) can be readily found by the knowledge of the vertex function at tree level, and is given by
\be
\Gamma^{\mu}(p,p-q) = \int d\omega \rho(\omega) \frac{i}{\slashed{p}-\omega + i\epsilon} i\gamma^{\mu} \frac{i}{\slashed{p}-\slashed{q}-\omega + i\epsilon}, \label{VecVertex1}
\ee 
up to transverse terms. This can be easily checked by replacing Eq. (\ref{VecVertex1}) into the left side of Eq. (\ref{VecWT}). Two-point functions can be obtained from the unamputed vertex by closing the fermions lines in Eq. (\ref{VecVertex1}) with the appropriate coupling \cite{Delbourgo1979,Arriola2003}. For example, the vacuum polarization tensor in the (1+1)D Schwinger model will be given by
\be
\Pi^{\mu \nu} = \int d\omega \rho(\omega) \int \frac{d^2 p}{(2\pi)^2} Tr\left\{ \frac{i}{\slashed{p}-\omega} i \gamma^{\mu} \frac{i}{(\slashed{p} - \slashed{q})-\omega} i \gamma^\mu \right\}. \label{Pimunu2}
\ee
The momentum integral appearing in the integrand on $\omega$ in Eq. (\ref{Pimunu2}) is the same integral we have already solved in the perturbative approach resulting in Eq. (\ref{VPTfinal}), with the replacement $m \rightarrow \omega$. It is an interesting feature of the spectral regularization the fact that the model, due to the presence of the spectral mass $\omega$, becomes naturally infrared regularized, even in a massless model. Our goal now will be to find a particular solution for $\rho(\omega)$ that fulfills the requirements we expect for the VPT of the Schwinger model, namely, for massless fermions: (i) to result in a finite and unambiguous amplitude; (ii) to be gauge invariant; (iii) to preserve the trace identity in Minkowski space of the amplitude. In addition to these requirements, we must expect the unitarity of the spectral density, i.e., (iv):
\be
\int d\omega \rho(\omega) = 1. \label{condition1}
\ee
First, this is an important condition that ensures that no constant ($\omega$ independent) content of the amplitudes will be modified by the spectral regularization. Besides, in the limit of large momentum, it implies the normalization of the fermion propagator, Eq. (\ref{LehmannProp}), to $i/\slashed{p}$ \cite{Arriola2003}.

It is important to observe that the spectral density can be used to implement different usual regularization schemes. To begin, it is evident that a spectral density $\rho(\omega)=\delta(\omega-m)$ reproduces the Feynman propagator for the fermion, from Eq. (\ref{LehmannProp}), and the non-regularized perturbative result for the VPT, from Eq. (\ref{Pimunu2}). The Pauli-Villars (PV) regularization scheme \cite{Pauli1949,Xue1991} can be implemented by choosing $\rho(\omega) = \sum_i C_i \delta(\omega - M_i)$, where $C_i$ and $M_i$ are the couplings and masses of the auxiliary fields introduced in the PV regularization. However, to implement regularizations that act directly in the integration domain of the momentum variable (sharp cutoff, proper time, Gaussian regularization, etc.) we must extend the spectral density to be a function also of the internal momentum. For example, in the sharp cut-off regularization scheme, we must have $\rho(\omega, p^2) = \delta(\omega - m) \Theta(\Lambda^2 - p^2)$, where $\Theta(x)$ is the Heaviside step function and $\Lambda$ is the covariant cut-off. 

Let us define some physical quantities within the spectral regularization. The fermion propagator is given by Eq. (\ref{LehmannProp}). It can be written as
\be
S(p)= \left(\int d\omega \rho(\omega) \frac{i}{p^2-\omega^2}\right) \slashed{p} + \left(\int d\omega \rho(\omega) \frac{i\omega}{p^2-\omega^2}\right) = A(p^2)\slashed{p} + B(p^2). \label{Ap_B}
\ee
It also can be described in terms of the fermion running mass $m(p^2)$ and field renormalization constant $Z(p)$ as
\be
S(p) = Z(p) i \frac{\slashed{p} + m(p^2)}{p^2 - m(p^2)^2},\label{FermionProp}
\ee
from which, by comparing with Eq. (\ref{Ap_B}), we obtain for the fermion mass
\be
m(p^2) = \frac{B(p^2)}{A(p^2)}.\label{mp2}
\ee
Notice that $m(p^2)$ is not the same mass as $m$, appearing in the Lagrangian, Eq. (\ref{lag}), for which $m \equiv 0$. $m(p^2)$, instead, is the dynamical effective mass computed from the non-perturbative Lehmann propagator, Eq. (\ref{LehmannProp}) \cite{Haeri1988}. The physical fermion mass is given by the pole of the fermion propagator, Eq. (\ref{FermionProp}), i.e., $p^2-m(p^2)^2=0$. The condition for massless fermions (i.e., the Schwinger model), necessary to ensure a solvable model and chiral symmetry (in the chiral Schwinger model), is the presence of a pole in the propagator at $p^2=0$, i.e., $m(0) = 0$, which implies, from Eq. (\ref{mp2}), in
\be
B(p^2=0)= -i \int d\omega \frac{\rho(\omega)}{\omega} = 0. \label{condition2}
\ee

A second relevant quantity to be taken into account in our analysis is the running coupling constant $e(p^2)$ or its spectral counterpart, $e(\omega)$. It considers the non-perturbative corrections to the electromagnetic coupling, i.e., the dependence of the coupling between the fermions and the gauge field with the spectral mass (as an energy scale). Within the spectral quark model \cite{Arriola2003}, the dependence of the couplings appearing in the vertexes functions showed up to be relevant for the determination of the spectral version of Goldberger-Treiman relation \cite{Goldberger1958} between the pion and quarks coupling and the pion electroweak decay constant. It also appears as a fundamental ingredient to the correct description of the axial anomaly within the model \cite{Ferreira2009}. Here, we will show that considering a spectral mass dependence for the electromagnetic coupling is necessary for the complete fulfillment of the expected requirements of the model. In this context, due to the dependence of the electromagnetic coupling with the spectral mass, we cannot factorize out this coupling in the vector Ward-Takahashi identity anymore, as we did in Eq. (\ref{VecWT}). Thus, the Ward-Takahashi identity should read
\be
q_{\mu} \Gamma_e^{\mu}(p,p-q) = e^2(p^2)S(p) - e^2((p-q)^2)S(p-q), \label{VecWT2}
\ee
where $e(p^2)$ is the running coupling constant and $\Gamma_e^{\mu}$ is the vector vertex function with the spectral coupling included, i.e.,
\be
\Gamma^{\mu}_{e}(p,p-q) = \int d\omega \rho(\omega) e^2(\omega) \frac{i}{\slashed{p}-\omega + i\epsilon} i\gamma^{\mu} \frac{i}{\slashed{p}-\slashed{q}-\omega + i\epsilon}. \label{VecVertex2}
\ee

The VPT can be written now, by using Eq. (\ref{VPTfinal}) and Eq. (\ref{Pimunu2}), as
\begin{eqnarray}
&&\Pi^{\mu \nu}(q^2) = \left( 2 \Delta^{\mu \nu} + \left(\frac{q^{\mu} q^{\nu}}{q^2} - \frac{g^{\mu \nu}}{2} \right) \frac{i}{\pi}\right) \oint_C \rho(\omega) e^2(\omega^2) d\omega \nonumber \\
&& - \left(\frac{q^{\mu} q^{\nu}}{q^2} - g^{\mu \nu} \right) \frac{i}{\pi} \oint_C \rho(\omega) e^2(\omega^2) \omega^2 Z_0(q^2,\omega^2) d\omega \nonumber \\
&& = \left( 2 \Delta^{\mu \nu} + \left(\frac{q^{\mu} q^{\nu}}{q^2} - \frac{g^{\mu \nu}}{2} \right) \frac{i}{\pi}\right) \rho_e - \left(\frac{q^{\mu} q^{\nu}}{q^2} - g^{\mu \nu} \right) \frac{i}{\pi} \rho_{Z0} \label{Pimunu3}
\end{eqnarray}
where $Z_0(q^2,\omega^2)$ is given by Eq. (\ref{Z0}) and we have defined
\be
\rho_e = \oint_C \rho(\omega) e^2(\omega^2) d\omega \label{rhoe}
\ee
and 
\be
\rho_{Z0} = \oint_C \rho(\omega) e^2(\omega^2) \omega^2 Z_0(q^2,\omega^2) d\omega.
\ee

The trace of the VPT in Minkowski space can now be obtained directly in (1+1)D from Eqs. (\ref{Pimumu}) and (\ref{Pimunu3}) (with the proper inclusion of the spectral mass, spectral density, and spectral coupling), giving, for $\Pi^{\mu}_{\mu}$,
\be
4 \oint_C \rho(\omega) e^2(\omega^2) \int \frac{d^2 p}{(2\pi)^2} \frac{\omega^2 }{(p^2-\omega^2)((p-q)^2-\omega^2)} = 2 \Delta^{\mu}_{\mu} \rho_e + \frac{i}{\pi} \rho_{Z0}. \label{Pimunu4}
\ee
The l.h.s. of Eq. (\ref{Pimunu4}) is convergent, can be explicitly computed, and results exactly equal to $\frac{i}{\pi} \rho_{Z0}$, i.e.,
\be
 \frac{i}{\pi} \rho_{Z0} = 2 \Delta^{\mu}_{\mu} \rho_e + \frac{i}{\pi} \rho_{Z0}, \label{TraceId}
\ee
showing that, to preserve the trace identity in exact (1+1)D, the condition to be satisfied is $\Delta^{\mu}_{\mu} = 0$ or $\rho_e = 0$. 

Now, to obtain a gauge invariant result from Eq. (\ref{Pimunu3}) we can assign a specific non-null value to $\Delta^{\mu \nu}$ in such a way that the first term in the r.h.s. of Eq. (\ref{Pimunu3}) (the coefficient of $\rho_e$) becomes gauge invariant, as discussed before, or, instead, we can assume $\rho_e = 0$, eliminating both the $\Delta^{\mu \nu}$ term and the non-gauge invariant $\left(\frac{q^{\mu} q^{\nu}}{q^2} - \frac{g^{\mu \nu}}{2}\right)$ term, preserving also the trace identity, Eq. (\ref{TraceId}).  So, to simultaneously satisfy the gauge invariance and the trace identity, obtaining a non-ambiguous result as a consequence, we must have $\rho_e = 0$. In other words, while the condition $\Delta^{\mu}_{\mu} = 0$ preserves the trace identity (eq. (\ref{TraceId})), it is not sufficient to ensure gauge invariance (eq. (\ref{Pimunu3})). However, $\rho_e = 0$ is a necessary and sufficient condition to preserve both gauge invariance and trace identity.

In this case, for $\rho_e = 0$, in order to avoid the trivial $\Pi^{\mu \nu} = 0$ solution, we also must have $\rho_{Z0} \ne 0$. That's why in the perturbative approach ($\rho(\omega) = \delta(\omega - m)$) there is no solution that simultaneously preserves the trace and the gauge identities, since, in this approach, $\rho_{Z0} = 0$ for $m=0$.

In summary, we have the following conditions to be fulfilled by the spectral function: 

i) Unitarity of the spectral function (normalization of the fermion propagator in the large momentum limit and preserving the constant content of the amplitudes), 
\be
\rho_0 \equiv \oint_C \rho(\omega) d\omega = 1; \label{rho0}
\ee
ii) Fermions with zero mass, 
\be
i B(p^2=0)= \int d\omega \frac{\rho(\omega)}{\omega} = 0; \label{Bp2}
\ee
iii) Non-ambiguous, gauge invariant, and trace identity preserving VPT:
\be
\rho_e = \oint_C \rho(\omega) e^2(\omega^2) d\omega = 0; \label{rho_e}
\ee
iv) Non-trivial VPT:
\be
\rho_{Z0} = \oint_C \rho(\omega) e^2(\omega^2) \omega^2 Z_0(q^2,\omega^2) d\omega \ne 0. \label{rhoZ0}
\ee
Conditions (i) to (iv) constitute a set of requirements to be fulfilled by any regularization procedure to be applied in the Schwinger model to simultaneously satisfy the physical (and mathematical) previously mentioned properties of the model in exact (1+1)D in a consistent way. In this sense, we can see that several of the usually employed regularization schemes do not satisfy these requirements, for instance: in sharp cutoff, with $\rho(\omega) = \Theta(p^2-\Lambda^2) \delta(\omega)$ we obtain $\rho_{Z0} = 0$ and $\rho_e = e^2(0)$, violating condition (iii); in the Pauli-Villars regularization scheme, with $\rho(\omega) = \sum_{i=0}^N C_i \delta(\omega - M_i)$, with $M_0=0$, $C_0 = 1$ and the connection limit is taken with $M_i \rightarrow \infty$ for $i \ne 0$ \cite{Morais2011}, we can have $\rho_e = 0$ and $\rho_{Z0} \ne 0$, satisfying (iii) and (iv) we if choose $\sum_i^N C_i = 0$ \cite{Morais2011}, and condition (ii) in the connection limit. However, we get $\rho_0 = 0$, i.e., condition (i) is not fulfilled. In the next section, we will show that we can find a particular solution for $\rho(\omega)$ and $e^2(\omega^2)$ that simultaneously satisfies conditions (i) to (iv), preserving thus gauge invariance and trace identity of the VPT in the (1+1)D Schwinger model.

\section{A Particular Solution}\label{AParticularSolution}

As already mentioned, the gauge technique consists in finding a particular solution for the vertex functions of the model, up to transverse parts, that satisfies all requirements of the model, in particular the Ward-Takahashi identity, Eq. (\ref{VecWT2}). In our approach, it means to find a particular representation for $\rho(\omega)$ and $e^2(\omega^2)$ in Eq. (\ref{VecVertex2}) that fulfills conditions (i) to (iv). In the spirit of the spectral regularization \cite{Arriola2003}, we also have to choose an appropriate complex contour $C$ that becomes, thus, part of the regularization.

We remark that the only energy scale present in two-point functions such as the VPT, computed from the vector vertex function solution, Eq. (\ref{VecVertex2}), is the external momentum squared norm, $q^2$, since fermions are massless in the Schwinger model. Thus, we must expect the spectral function $\rho(\omega)$ and spectral coupling $e^2(\omega^2)$ to be implicit functions of this only scale, at least within the two-point functions. We will show further that this scale-dependent spectral function does not jeopardize conditions (i) and (ii), which must be independent of $q^2$. 

We observe, from Eq. (\ref{rhoZ0}), that if $\rho(\omega) e^2(\omega^2) = \frac{f(\omega)}{\sqrt{q^2-4\omega^2}}$ for a non-singular arbitrary $f(\omega)$, we got two poles on the integrand of Eq.(\ref{rhoZ0}) without introducing any pole on the integrand of $\rho_e$, Eq. (\ref{rho_e}). Together with shifting the poles to the upper half of the $\omega$ complex plane of an infinitesimal amount $\epsilon$ and integrating along a closed path $C$ involving both poles, this guarantees the conditions (iii) and (iv) to be satisfied. So, considering also conditions Eqs. (\ref{rho0}) and (\ref{Bp2}),  a particular solution for conditions (i) to (iv), up to transverse terms and constant multiplicative factors, is given by
\be
\rho(\omega)  = \frac{2i}{\pi} \omega \frac{e^{-|1-\frac{4\omega^2}{q^2}|}}{q^2 - 4 \omega^2} \label{rhoomega}
\ee
and
\be
e^2(\omega^2) = \frac{e_0^2}{\pi} \sqrt{1 - 4 \frac{\omega^2}{q^2}}, \label{e2omega}
\ee
to be integrated in the aforementioned complex contour $C$. The $e^{-|1-\frac{4\omega^2}{q^2}|}$ factor was included only to ensure that the integrand vanishes at the $|\omega| \rightarrow \infty$ limit.

Conditions (i)--(iv), i.e., Eq. (\ref{rho0})--(\ref{rhoZ0}) can be directly verified. In fact, the product $\rho(\omega) e^2(\omega^2)$ has no pole inside the complex path $C$, so $\rho_e = \oint_C \rho(\omega) e^2(\omega^2) d\omega = 0$, Eq. (\ref{rho_e}) is satisfied. The integrand of $\rho_0$, i.e., $\rho(\omega)$, presents two single poles located at $\omega = \pm \frac{q}{2}$, with $q=\sqrt{q^2}$. The residues at these poles are both equal to $-\frac{i}{4\pi}$, and considering the displacement of the poles from an infinitesimal amount $i\epsilon$ into the contour $C$, we get Eq. (\ref{rho0}).

The integrand of Eq. (\ref{Bp2}) has two single poles at $\omega=\pm \frac{q}{2}$. We get two pairs of residues with opposite signs, and thus the integral $B(p^2)$, Eq. (\ref{Bp2}), vanishes. 

Finally, we are left with $\rho_{Z0}$ to be solved. Replacing Eqs. (\ref{Z0}), (\ref{rhoomega}) and (\ref{e2omega}) in Eq. (\ref{rhoZ0}), we get
\be
\rho_{Z0}(q^2) = \frac{-8 i e_0^2}{\pi^2} \oint_C \frac{ \omega^3 e^{-|1-4\omega^2/q^2}|}{q^2 (q^2 - 4\omega^2)} \textrm{ arctanh}\left( \frac{1}{\sqrt{1-4 \omega^2/q^2}} \right) d\omega.
\ee
Its integrand has two poles located at $\omega = \pm \frac{q}{2}$. At these poles,
\be
\lim_{\omega \rightarrow \pm q/2} \textrm{arctanh}\left(\frac{1}{\sqrt{1-4 \omega^2/q^2}}\right) = \frac{\pi}{2}. 
\ee
Performing the integration by residues in the contour C, we obtain
\be
\rho_{Z0}(q) = -i e_0^2,
\ee
retrieving the exact conventional result for the vacuum polarization tensor in the Schwinger model, 
\be
\Pi^{\mu \nu}(q) = -\frac{e_0^2}{\pi} \left( \frac{q^\mu q^\nu}{q^2} - g^{\mu \nu} \right).
\ee

This concludes our demonstration that the particular solution for the spectral density, Eq. (\ref{rhoomega}), and for the spectral coupling, Eq.(\ref{e2omega}), satisfy the spectral conditions given by Eqs. (\ref{rho0}) to (\ref{rhoZ0}), ensuring gauge invariance and mathematical consistency (preservation of the trace identity) within the Schwinger model in exact (1+1)D.

For completeness, let us show, as previously mentioned, that although $\rho(\omega)$ and $e^2(\omega^2)$ are implicitly dependent on the external energy scale $q^2$, it does not jeopardize the scale independent relations given by Eqs. (\ref{rho0}) and (\ref{Bp2}).

From Eq. (\ref{rho0}), the unitarity condition is given by
\be
\rho_0 = \oint_C \rho(\omega) d\omega = \oint_C \frac{2i}{\pi} \omega \frac{e^{-|1-\frac{4\omega^2}{q^2}|}}{q^2 - 4 \omega^2} d\omega=1.\label{unitcond}
\ee
The result of Eq.(\ref{unitcond}) is in fact independent on $q^2$, since, by performing the change of variable a$x=\frac{\omega}{q}$, with $q = \sqrt{q^2}$, as before, we get
\be
\rho_0 = \oint_C \frac{2i}{\pi} x \frac{e^{-|1-x^2|}}{1-x^2} dx,\label{unitcond2}
\ee
with the poles located at $x=\pm 1$. So, neither the integrand nor the poles of Eq.(\ref{unitcond2}) are scale-dependent. The same feature can be observed also in Eq. (\ref{rho_e}) with the same change of variable. For $B(p^2)$, Eq. (\ref{Bp2}), this independence on the external energy scale $q^2$ is not observed, since, by changing $\omega \rightarrow q x$, we get two of the poles at $x = \pm p/q$. However, as the poles have opposite signs, they cancel out in the final result, i.e., $B(p^2) = 0$ for any value of $q$.

\section{The Chiral Schwinger Model}
Now let us analyze the consequences of the employment of the spectral regularization on the chiral Schwinger model. The chiral Schwinger model \cite{Rajamaran1985}, a U(1) gauge field coupled to massless chiral fermions in (1+1)D, is defined by the Lagrangian density
\be
\mathcal{L}= \bar{\psi} (i \slashed{\partial} + m) \psi - e_0 \bar{\psi} \gamma_\mu (1+\gamma^5) A^{\mu} \psi, \label{lagCS}
\ee
where $\gamma^5=\gamma^0 \gamma^1$. The quantity of interest, in this case, is the vector/axial-vector two-point function for the gauge boson, due to its similarity to the Quantum Chromodynamics' Adler-Bardeen-Bell-Jackiw anomaly. In fact, as we shall see also in our approach to the problem, it is impossible to obtain a gauge-invariant result, expressed, in the vector/axial-vector two-point function, as the impossibility of simultaneously satisfying the vector and axial vector Ward identities.

The vector/axial-vector two-point function for the gauge boson is given by
\be
\Pi_{VA}^{\mu \nu}(q) = \oint_C d\omega \rho(\omega) e^2(\omega^2) \int \frac{d^2 p}{(2\pi)^2} Tr\left\{ i \gamma^\mu \frac{i}{\slashed{p}-\omega} i \gamma^{\nu} \gamma^5 \frac{i}{(\slashed{p} - \slashed{q})-\omega} \right\}, \label{TPVA}
\ee
where we have already introduced the spectral regularization. At this point, it is important to notice that, as already mentioned in the Introduction, the $\gamma^5$ algebra can bring serious problems to the correct treatment of chiral models. We mention, again, that there is no natural extension of the $\gamma^5$ object to non-integer dimensions. Besides, it was shown by Viglioni et all \cite{Viglioni2016} that the $\gamma_5$ algebra is ambiguous inside divergent integrals, in both (1+1)D and (3+1)D. The spectral regularization, however, has the feature of, by the appropriate choice of spectral conditions, to eliminate any divergent and/or ambiguous integral in the computation of quantum amplitudes, as we saw in the previous sections. Only convergent non-ambiguous results are left, and in this case, we can employ any relation obtained from the exact (1+1)D $\gamma^5$ algebra. So, let us compute the results and show how the spectral regularization eliminates the ambiguous results potentially appearing in the calculation of $\Pi_{VA}^{\mu \nu}(q)$.

We use the following $\gamma^5$ relations, easily demonstrated by using any explicit representation of the $\gamma$ matrices in (1+1)D:
\be
Tr\{\gamma^\mu \gamma^\nu \gamma^5\}=-2\varepsilon^{\mu\nu},
\ee
\be
\gamma^\nu \gamma^5 = 2 \varepsilon^\nu_{\textrm{ }\alpha} \gamma^\alpha, \label{gammamugamma5}
\ee
and, as a consequence of Eq. (\ref{gammamugamma5}),
\be
Tr\{\gamma^\mu \gamma^\rho \gamma^\nu \gamma^5 \gamma^\sigma\} = 2 \varepsilon^\nu_{\textrm{ }\alpha} (g^{\mu\rho} g^{\alpha\sigma} - g^{\mu\alpha} g^{\rho\sigma} + g^{\mu\sigma}g^{\rho\alpha}).
\ee

Carrying out the calculations as before, we obtain
\begin{eqnarray}
&&\Pi_{VA}^{\mu \nu}(q)= 2 \oint_C d\omega \rho(\omega) e^2(\omega^2) \Bigg\{  \varepsilon^\nu_{\textrm{ }\alpha} \Delta^{\mu\alpha} +  \nonumber \\
&& \frac{i}{2\pi} \left(\varepsilon^\nu_{\textrm{ }\alpha} (2 q^{\mu} q^{\alpha} - q^2 g^{\mu \alpha}) \left(\frac{1}{q^2} - \frac{\omega^2}{q^2} Z_0 \right) -  \varepsilon^{\mu\nu} \omega^2 Z_0 \right)\Bigg\}, \label{PIVA} 
\end{eqnarray}
where $Z_0 = Z_0(q^2,\omega^2)$, given by Eq. (\ref{Z0}).

The non-regularized perturbative result for massless fermions can be retrieved by taking $\rho(\omega) = \delta(\omega)$, and we observe that, in such a case, the two last terms of Eq. (\ref{PIVA}) vanish, and the preservation (or violation) of the vector (or axial-vector) Ward identity can only be assured by a specific choice of the regularization ambiguous term $\Delta^{\mu\nu}$, as pointed out in \cite{Jackiw2000,Viglioni2016}. Let us, however, focus on the spectral regularized result. By using the definitions of the spectral quantities $\rho_e$ and $\rho_{Z0}$, Eqs. (\ref{rhoe}) and (\ref{rhoZ0}) respectively, we get
\be
\Pi_{VA}^{\mu \nu}(q)=2 \varepsilon^\nu_{\textrm{ }\alpha} \Delta^{\mu\alpha} \rho_e + 
\frac{i}{2\pi} \left\{\varepsilon^\nu_{\textrm{ }\alpha} (2 q^{\mu} q^{\alpha} - q^2 g^{\mu \alpha}) \left(\frac{\rho_e}{q^2} - \frac{\rho_{Z0}}{q^2}  \right) - \varepsilon^{\mu\nu} \rho_{Z0} \right\}. \label{PIVAfinal}
\ee

We can now compute the Ward identities relative to each (vector or axial-vector) vertex from the two-point function, Eq. (\ref{PIVAfinal}), obtaining
\be
q_\mu \Pi_{VA}^{\mu \nu}(q) = \left( 2 \varepsilon^\nu_{\textrm{ }\alpha} q_\mu \Delta^{\mu\alpha} + 
\frac{i}{2\pi} q^{\alpha} \varepsilon^\nu_{\textrm{ }\alpha} \right) \rho_e, \label{vecWI}
\ee
where the terms proportional to $\rho_{Z0}$ canceled out due to the antisymmetry of the tensor $\varepsilon^\nu_{\textrm{ }\alpha}$, and
\be
q_\nu \Pi_{VA}^{\mu \nu}(q) = 2 q_\nu \varepsilon^\nu_{\textrm{ }\alpha} \Delta^{\mu\alpha} \rho_e - 
\frac{i}{2\pi} \left(q_\nu \varepsilon^{\nu \mu} \rho_e - 2 q_\nu \varepsilon^{\nu \mu} \rho_{Z0} \right). \label{axvecWI}
\ee

It is interesting to observe that the same spectral condition employed to ensure gauge invariance of the Schwinger model, $\rho_e = 0$, is also sufficient to satisfy the vector Ward Identity, Eq. (\ref{vecWI}), giving $q_\mu \Pi_{VA}^{\mu \nu} = 0$. In such a case, we are left with the violation of the axial current conservation, i.e., $q_\nu \Pi_{VA}^{\mu \nu} = \frac{i}{\pi} q_\nu \varepsilon^{\nu \mu} \rho_{Z0}$. Besides, both Eqs. (\ref{vecWI}) and (\ref{axvecWI}) cannot simultaneously vanish unless $\rho_e = 0$ and $\rho_{Z0} = 0$, implying in the non-physical triviality of the vector/axial-vector two-point function, Eq. (\ref{PIVAfinal}). This is the well-known manifestation of the anomaly within the Chiral Schwinger model. It is worth noticing that, by choosing to preserve the vector Ward Identity, i.e., $\rho_e = 0$, both results from Eqs. (\ref{vecWI}) and (\ref{axvecWI}) became independent from the regularization ambiguous $\Delta^{\mu\nu}$ term.

However, as discussed in section \ref{TheModel}, in a covariant regularization in exact (1+1)D we have $\Delta^{\mu\nu} = 0$. In this case, if one chooses (or if the model imposes it) to preserve the axial-vector Ward Identity, violating the vector one, the spectral condition $\rho_e = 2 \rho_{Z0}$ can be employed, resulting in $q_\mu \Pi_{VA}^{\mu \nu} = \frac{i}{\pi} q_\mu \varepsilon^{\nu \mu} \rho_{Z0}$ and $q_\nu \Pi_{VA}^{\mu \nu} = 0$. Any other specific choice for $\rho_e \ne 0$ will result, of course, in the violation of both Ward Identities, and the expected freedom in which Ward Identity should be violated or preserved \cite{Rajamaran1985,Viglioni2016} is observed, even in the absence of the ambiguous $\Delta^{\mu\nu}$ term. This feature is also observed in the study of the QCD chiral anomaly within the spectral quark model \cite{Ferreira2009}, where the Ward identity to be preserved (or violated) is determined by the dependence of the axial coupling with the spectral mass.

Spectral conditions for the unitarity of the spectral function, Eq. (\ref{rho0}) and massless fermions, Eq. (\ref{Bp2}), together with the relation between $\rho_e$ and $\rho_{Z0}$ and non-triviality of the vector/axial-vector two-point function constitute the set of spectral conditions to guide the choice of the spectral density $\rho(\omega)$ and spectral coupling $e^2(\omega)$ for the chiral Schwinger model. The particular solution presented in section \ref{AParticularSolution}, as also in the Schwinger model approach, can be employed in order to preserve the vector Ward identity, violating the axial-vector one.

\section{Conclusion}

In this work, we have shown that the spectral regularization, a non perturbative regularization scheme previously applied to chiral quark models, can be successfully applied to the Schwinger model and chiral Schwinger model in exact (1+1)D in order to obtain mathematical consistent and ambiguities free results, in contrast with others previous perturbative results. The results are gauge invariant, in the Schwinger model, and correctly display the anomalous results in the chiral Schwinger model. The spectral regularization is carried out in exact integer dimensions, and eliminates divergences and ambiguities by means of a proper choice of the spectral conditions, so the spectral regularization also avoids the $\gamma_5$ algebra subtleness present in non-integer dimensions or within divergent integrals. In this sense, the use of the spectral regularization could be of fundamental importance in the study of physical systems where the Schwinger model (or other (1+1)D fermionic models) emerges as an effective model, avoiding the presence of ambiguities and lack of mathematical consistency in its treatment. In summary, our results reinforce the strength of the spectral regularization as a mathematical consistent, divergence and ambiguity free, and also symmetry/symmetry violation preserving regularization scheme. 

\section*{Acknowledgments}
The authors are grateful to FAPEMIG (Brazilian funding agency). This study was financed in part by the Coordena\c{c}\~{a}o de Aperfei\c{c}oamento de Pessoal de N\'{i}vel Superior - Brasil (CAPES) - Finance Code 001.

\end{document}